\documentclass{article}
\usepackage[utf8]{inputenc}
\usepackage{lipsum} % For placeholder text if needed
\usepackage{graphicx} % For figures
\usepackage{amsmath,amssymb} % For math environments
\usepackage{hyperref} % For clickable references/links
\usepackage{url}      % For URL typesetting
\usepackage{cite}     % For better citation handling
\usepackage[sort&compress,numbers]{natbib}
\usepackage{authblk}   % in the preamble
\usepackage{enumitem}

\title{Security‑First AI: Foundations for Robust and Trustworthy Systems}

\author[1,2]{Krti Tallam}
\affil[1]{Sentinel.AI, San Francisco CA, USA}
\affil[2]{EECS\\ University of California at Berkeley}
\affil[ ]{\texttt{ktallam@sentinel-security.ai}}

\date{\today}

\begin{document}

\maketitle

\begin{abstract}
The conversation around artificial intelligence (AI) often focuses on safety, transparency, accountability, alignment, and responsibility. However, AI security (i.e., the safeguarding of data, models, and pipelines from adversarial manipulation) underpins all of these efforts. This manuscript posits that \textbf{AI security} must be prioritized as a foundational layer. We present a hierarchical view of AI challenges, distinguishing security from safety, and argue for a security-first approach to enable trustworthy and resilient AI systems. We discuss core threat models, key attack vectors, and emerging defense mechanisms, concluding that a metric-driven approach to AI security is essential for robust AI safety, transparency, and accountability.
\end{abstract}

\section{Introduction}
The rapid advancement of artificial intelligence (AI) has sparked both excitement and significant concern. AI systems are now integral to critical societal sectors, including healthcare, finance, transportation, and national security, raising urgent questions about their safety, transparency, ethical alignment, accountability, and overall responsibility. Extensive research and policy discussions have emerged to ensure these technologies operate without causing unintended harm, generating opaque or biased outcomes, or deviating from human values.

While these considerations are vital, this manuscript asserts that \textbf{AI security} (the protection of data, models, and computational pipelines from adversarial manipulation and unauthorized access), must be prioritized as a foundational prerequisite. By integrating end-to-end security protocols throughout the AI lifecycle, from initial data collection to final model deployment, we substantially reduce the risk of catastrophic failures and ensure that subsequent safety and ethical mechanisms rest on a resilient infrastructure.

\subsection{Scope and Significance of AI Security}
The domain of AI security is expansive, covering multiple attack vectors and threat models. Notable examples include \textbf{data poisoning}, where adversaries inject or manipulate training data to alter model behavior at inference time~\cite{biggio2012poisoning}; \textbf{model inversion attacks}, which reconstruct sensitive training data from model outputs~\cite{fredrikson2015model}; and \textbf{adversarial examples}, carefully crafted perturbations designed to deceive otherwise accurate models~\cite{goodfellow2015explaining}. Additional concerns include \textbf{model extraction attacks}, in which attackers systematically query proprietary models to replicate them~\cite{tramer2016stealing}, and \textbf{membership inference attacks}, which aim to identify whether particular data points were included in training sets, thus compromising user privacy~\cite{shokri2017membership}.

Addressing these vulnerabilities is essential for several critical reasons:
\begin{itemize}
    \item \textbf{Preserving Trust:} Users and stakeholders are more likely to embrace AI technologies that consistently demonstrate resilience against adversarial threats and malicious manipulation.
    \item \textbf{Regulatory Compliance:} As global data protection regulations evolve, robust AI security practices are increasingly necessary to meet legal obligations and ethical expectations.
    \item \textbf{Protecting Intellectual Property:} Significant commercial and academic investments in AI models necessitate safeguarding proprietary techniques from unauthorized extraction or reverse-engineering.
    \item \textbf{Mitigating Risk in Critical Domains:} Ensuring AI-driven solutions in sectors such as healthcare, finance, and autonomous vehicles remain robust under adversarial conditions is paramount to preventing severe real-world consequences.
\end{itemize}

By elevating AI security as a foundational priority, this manuscript advocates for a proactive, defense-oriented paradigm that fortifies trust, ensures regulatory compliance, protects intellectual assets, and significantly reduces operational risks inherent in deploying AI systems at scale.

\subsection{Why Security Must Come First}
Prioritizing security from the outset is essential to safeguarding AI systems against both existing and emerging threats \cite{tallam2025proactive}. Modern AI, characterized by increasingly complex models, particularly deep neural networks, and massive datasets, introduces novel vulnerabilities exploitable by adversaries. Without rigorous security protocols, even well-designed AI safety measures and ethical frameworks become susceptible to compromise. For instance, an AI system intended to align closely with human values can still be manipulated via adversarially crafted inputs if its underlying infrastructure lacks robust security \cite{tallam2025immune}. Similarly, transparency initiatives intended to provide interpretability and accountability may unintentionally expose sensitive model information, facilitating more sophisticated and targeted adversarial attacks.

Embedding security early in the AI lifecycle significantly strengthens all subsequent safety and ethical initiatives. A comprehensive, security-first approach ensures that foundational vulnerabilities are addressed proactively, augmenting the effectiveness and reliability of transparency, accountability, and alignment strategies. Ultimately, this approach not only protects individual systems but also reinforces broader trust across the AI ecosystem, enabling safer deployments, ethical governance practices, and consistently reliable outcomes.

\subsection{Structure of This Manuscript}
This manuscript emphasizes the foundational role of AI security, presenting it as an essential precursor to broader AI safety measures and ethical considerations. The manuscript systematically guides readers from foundational concepts through practical defensive strategies and ultimately demonstrates the critical interplay between security and safety. The structure is organized as follows:

\begin{itemize}
    \item \textbf{Section~\ref{sec:background} (Background on AI Security):} 
    We first provide an in-depth review of AI security's current landscape. We examine widely recognized threat models and attack vectors, such as data poisoning, adversarial examples, model extraction, and membership inference. The section incorporates both theoretical insights and real-world case studies, underscoring the diverse adversarial threats facing contemporary AI systems and reviewing key literature shaping this rapidly evolving field.

    \item \textbf{Section~\ref{sec:hierarchy} (Hierarchy of AI Challenges):} 
    Next, we clarify the distinction between \textbf{AI security} and \textbf{AI safety} through a hierarchical framework. While both concepts aim to prevent harm, they target different operational layers within the AI lifecycle. This section precisely defines each term and illustrates, through detailed case studies, how inadequate security measures can undermine broader safety objectives, thereby highlighting the necessity of an integrated, layered security approach.

    \item \textbf{Section~\ref{sec:security-first} (The Security-First Approach):} 
    This section explores the rationale for prioritizing a security-first perspective. We present concrete examples of how adversarial vulnerabilities can severely impact AI systems designed with robust safety and ethical frameworks. Additionally, we discuss organizational and policy implications, including resource allocation strategies, specialized security expertise, and the critical importance of ongoing security audits to maintain robust system defenses.

    \item \textbf{Section~\ref{sec:defenses} (Defensive Techniques and Best Practices):} 
    We then delve into state-of-the-art defensive strategies, including adversarial training, differential privacy, robust model architectures, and secure deployment procedures. Practical guidance on securing the AI pipeline -- from data collection and model training to real-world inference -- highlights specific best practices such as comprehensive logging, continuous monitoring, and incident response protocols, illustrating how theoretical principles translate into effective real-world security.

    \item \textbf{Section~\ref{sec:integration} (Integrating Security with Safety):} 
    Building upon preceding discussions, this section demonstrates how AI security and AI safety frameworks can be seamlessly integrated. We argue that effective safety measures depend fundamentally on secure infrastructures, providing illustrative examples from industry and academia where organizations successfully balance robust security with ethical considerations, thus promoting greater stakeholder trust in AI technologies.
\end{itemize}

Collectively, these sections form a comprehensive framework emphasizing the need to proactively prioritize AI security. By securing foundational infrastructure, subsequent safety, transparency, and accountability initiatives can fulfill their objectives more effectively. As AI continues to evolve, maintaining this security-first perspective is imperative for developing intelligent systems that are not only useful but consistently trustworthy and resilient against adversarial threats.

\section{Background and Motivation}\label{sec:background}

While considerable attention has been devoted to AI safety and ethical alignment, comparatively fewer studies explicitly focus on \textbf{AI security} as an independent and critical research domain. In this manuscript, we define AI security broadly, encompassing a range of technical measures and organizational practices dedicated to protecting data, models, and computational pipelines from adversarial manipulation and unauthorized access. Without robust and proactive security measures, even the most rigorously designed safety protocols and ethical frameworks may be compromised by adversaries.

Recent advances in adversarial machine learning have exposed severe vulnerabilities within modern AI systems. For instance, \textbf{data poisoning attacks} illustrate how strategically injected malicious data during training can degrade or even completely subvert model performance~\cite{biggio2012poisoning, steinhardt2017certified}. Such attacks highlight the urgent need for ensuring data integrity, as compromised training datasets can lead to unpredictable and hazardous outcomes in deployed systems.

Similarly concerning are \textbf{model inversion attacks}, which exploit model outputs to reconstruct sensitive information about training datasets~\cite{fredrikson2015model, melis2019exploiting}. These breaches pose significant risks to user privacy and introduce regulatory compliance challenges, particularly within highly regulated sectors such as healthcare and finance.

A particularly prominent vulnerability arises from \textbf{adversarial examples}, carefully engineered inputs containing subtle, often imperceptible perturbations that reliably deceive high-performing models~\cite{goodfellow2015explaining, carlini2017towards, kurakin2016adversarial}. The susceptibility of state-of-the-art AI to such minimal alterations raises serious questions about their robustness and suitability for safety-critical applications.

Additionally, \textbf{model extraction attacks} exacerbate these vulnerabilities by allowing attackers to replicate proprietary AI models through systematic querying and analysis~\cite{tramer2016stealing}. This practice not only threatens intellectual property rights but also facilitates further adversarial activities, such as the targeted crafting of adversarial examples or the cloning of competitive models.

Another critical concern arises from \textbf{membership inference attacks}, where adversaries attempt to determine if specific data instances were used during model training~\cite{shokri2017membership, yeom2018privacy}. Such attacks jeopardize individual privacy, creating significant risks under stringent data protection regulations, such as GDPR or HIPAA.

Importantly, these threats are not merely theoretical concerns; they have substantial practical implications across diverse sectors. For example, adversarial attacks in healthcare can mislead diagnostic AI systems, potentially resulting in misdiagnoses and compromised patient safety~\cite{finlayson2019adversarial}. In finance, similar adversarial manipulations can distort risk assessment models, influencing investment decisions and potentially destabilizing financial markets~\cite{bhattacharyya2011data}. Autonomous vehicles also remain vulnerable, as empirical studies demonstrate that adversarial inputs can significantly degrade object detection and navigational capabilities, raising serious safety concerns~\cite{huval2015empirical}.

The increasingly distributed nature of AI infrastructure across cloud, edge, and hybrid environments further expands potential attack surfaces. Modern AI deployments often comprise heterogeneous ecosystems, where security breaches in a single component can propagate throughout the system~\cite{zhang2018securing}. Such complexity demands comprehensive and integrated security approaches that span the entire AI lifecycle, from initial data acquisition and model training through to deployment and continuous operational monitoring \cite{tallam2025odi}.

Thus, the imperative for prioritizing AI security emerges from several core motivations:

\begin{itemize}
    \item \textbf{Protecting Sensitive Data:} Preventing external and internal threats from compromising data integrity and confidentiality.
    \item \textbf{Maintaining Model Integrity:} Ensuring model parameters and decision-making processes remain secure against adversarial manipulations that can cause incorrect predictions or system-wide failures.
    \item \textbf{Ensuring Regulatory Compliance:} Adhering proactively to evolving global data protection and privacy regulations.
    \item \textbf{Safeguarding Intellectual Property:} Preventing unauthorized extraction, duplication, or reverse-engineering of proprietary AI models and methodologies.
    \item \textbf{Refining System Robustness:} Building resilient systems capable of operating reliably even under adversarial conditions, thereby reinforcing trust among users and stakeholders.
\end{itemize}

\noindent
In light of these motivations, this manuscript advocates a security-first paradigm as a fundamental precursor to broader AI safety and ethical frameworks. By proactively addressing foundational vulnerabilities, we can establish resilient AI systems capable of delivering reliable performance under benign circumstances while preserving their integrity in the presence of adversarial threats.

\subsection{Threat Models in AI}
A \textbf{threat model} outlines the assumptions about an adversary’s capabilities, goals, and the level of access they possess to a system. In AI, especially in the context of adversarial machine learning, establishing a clear threat model is critical because it guides the design of effective defense mechanisms. Broadly, threat models in AI can be classified into three categories:

\begin{itemize}
    \item \textbf{White-box Threat Model:}  
    Under this model, the adversary is assumed to have full access to the target model’s internal details, including its architecture, parameters, training data, and any preprocessing steps. This model represents a worst-case scenario where attackers can leverage complete information to craft precise adversarial examples \cite{goodfellow2015explaining, carlini2017towards}. Evaluations in the white-box setting often serve as stress tests, demonstrating how robust, or fragile, a defense is when facing an all-knowing adversary.

    \item \textbf{Gray-box Threat Model:}  
    In a gray-box scenario, the adversary has limited or partial knowledge about the model. This might include knowing the model’s architecture but not its exact parameters, or having access to a subset of the training data without full insight into the training process \cite{papernot2016limitations}. Gray-box models reflect more realistic situations where certain information might be exposed (for example, through open-source implementations or shared benchmarks), yet complete transparency is not available. Defenses in this setting need to account for an adversary that can exploit partial knowledge while still operating under uncertainty.

    \item \textbf{Black-box Threat Model:}  
    Here, the adversary has no access to the internal workings of the model and can only interact with it through its input-output behavior. Despite these limitations, attackers can still generate adversarial examples by observing model responses, using techniques such as query-based attacks or leveraging the transferability of adversarial examples from surrogate models \cite{tramer2016stealing, papernot2017practical}. This model is particularly relevant for deployed systems, such as commercial APIs or closed-source products, where internal details are guarded but the system remains vulnerable to external probing.
\end{itemize}

Understanding the relevant threat model(s) for a given application is a prerequisite to designing robust defenses. A comprehensive threat model analysis enables researchers and practitioners to tailor security strategies effectively, as defenses that are strong in a white-box setting might not necessarily translate to robustness in a black-box environment. Moreover, considering multiple threat models in tandem allows for the development of adaptive, multi-layered defense mechanisms that better address the diverse risks encountered in real-world AI applications.

\subsection{Common Attack Vectors}
AI systems, while powerful, are susceptible to a variety of attacks that exploit weaknesses at different stages of the AI lifecycle. Understanding these vectors is essential for developing robust defenses. Below are some of the primary attack vectors:

\begin{itemize}
    \item \textbf{Data Poisoning:}  
    Data poisoning attacks involve the deliberate injection or manipulation of training data to corrupt the learning process. Attackers can introduce subtle modifications to a subset of the training data, causing the model to learn incorrect patterns or behaviors. This type of attack can be particularly insidious because the corrupted data may appear innocuous to human reviewers \cite{biggio2012poisoning}. Recent studies have extended these ideas to scenarios where poisoning is used to introduce backdoors or trigger specific responses in the deployed model, even when only a small portion of the data is compromised.

    \item \textbf{Model Inversion Attacks:}  
    In model inversion attacks, adversaries exploit the outputs of a machine learning model to infer sensitive information about the training data. By carefully analyzing the model's predictions, an attacker can reconstruct input features that are correlated with private attributes, potentially revealing confidential or personal data \cite{fredrikson2015model}. Such attacks pose significant privacy risks, especially when models are trained on sensitive information such as medical records or personal identification data.

    \item \textbf{Adversarial Examples:}  
    Adversarial examples are inputs to machine learning models that have been intentionally perturbed in a way that is almost imperceptible to humans but leads the model to make incorrect predictions. These carefully crafted perturbations can cause even state-of-the-art models to fail, raising concerns about the reliability and safety of AI systems in critical applications \cite{goodfellow2015explaining}. The development of adversarial examples has spurred extensive research into robust training techniques and defensive strategies to mitigate these vulnerabilities.

    \item \textbf{Model Extraction:}  
    Model extraction attacks involve systematically querying a target model to reconstruct its underlying parameters or to build a surrogate model that mimics its behavior. Attackers can use this surrogate model to understand and exploit the original model's decision boundaries, effectively bypassing proprietary protections and intellectual property rights \cite{tramer2016stealing}. This form of attack not only endangers commercial interests but also increases the risk of deploying cloned models that may be more vulnerable to subsequent adversarial manipulation.

    \item \textbf{Membership Inference:}  
    Membership inference attacks aim to determine whether a specific data point was part of a model’s training dataset. By exploiting subtle differences in the model’s responses to inputs that were seen during training versus those that were not, adversaries can infer membership status \cite{shokri2017membership}. Such attacks have serious implications for privacy, as they can be used to reveal sensitive information about individuals, particularly in contexts where training data includes confidential or personally identifiable information.
\end{itemize}

Each of these attack vectors exposes unique vulnerabilities within AI systems, emphasizing the need for a multi-layered security strategy. Addressing these threats requires not only improved detection and defense mechanisms but also a fundamental rethinking of how data, models, and operational processes are secured throughout the AI lifecycle.

\section{The Hierarchy of AI Challenges}\label{sec:hierarchy}

In the complex ecosystem of AI research and deployment, it is crucial to disentangle the different challenges that must be addressed to build robust systems. Two of the most critical, yet sometimes conflated, domains are \textbf{AI security} and \textbf{AI safety}. In this section, we define these concepts, outline their unique roles, and explain how they interact within the broader landscape of AI challenges.

\subsection{Defining AI Security}
\textbf{AI security} is concerned with protecting the core components of an AI system -- data, models, and the operational pipelines -- from adversarial attacks and unauthorized access. Its primary objective is to ensure that the system remains robust in the face of both deliberate and opportunistic threats. Key facets of AI security include:
\begin{itemize}
    \item \textbf{Developing Robust Metrics:} Establishing quantitative measures that can assess the vulnerabilities of data sources, model architectures, and deployment pipelines. For example, vulnerability scoring systems and anomaly detection metrics have been proposed to monitor and quantify security risks in real time \cite{papernot2018deep, steinhardt2017certified}.
    \item \textbf{Monitoring Data Integrity and Model Behavior:} Continuously tracking the inputs, intermediate states, and outputs of AI systems to detect anomalous behavior that may indicate an ongoing attack \cite{finlayson2019adversarial, chen2018detecting}. 
    \item \textbf{Implementing Defensive Techniques:} Deploying strategies such as adversarial training, defensive distillation, and differential privacy to harden models against specific attack vectors \cite{goodfellow2015explaining, papernot2016distillation, abadi2016deep}.
    \item \textbf{Continual Assessment of New and Evolving Threats:} Given the rapid evolution of attack methodologies, AI security requires dynamic adaptation and regular reassessment of risk models. This includes staying abreast of emerging research and adapting defense mechanisms accordingly \cite{carlini2017towards, tramer2017ensemble}.
\end{itemize}
AI security, therefore, is not a one-time installation of safeguards but a continuous process of evaluation, adaptation, and improvement—a necessary foundation for the reliable deployment of AI systems.

\subsection{Distinguishing AI Safety}
While AI security deals with protecting the system from external attacks and internal corruptions, \textbf{AI safety} focuses on ensuring that AI systems operate as intended and do not inadvertently cause harm. Its emphasis lies on:
\begin{itemize}
    \item \textbf{Ensuring Transparency in Decision-Making:} Developing methods that allow stakeholders to understand and interpret how and why a system arrives at its decisions, thereby building trust and facilitating oversight \cite{doshi2017towards}.
    \item \textbf{Establishing Frameworks for Accountability and Ethical Use:} Crafting policies and technical mechanisms that ensure decision-making is aligned with ethical standards and legal requirements, and that mechanisms for redress exist when failures occur \cite{mittelstadt2016ethics, jobin2019global}.
    \item \textbf{Aligning AI Objectives with Human Values:} Embedding human-centered design principles into the system so that the behavior of the AI remains consistent with societal norms and expectations \cite{bostrom2014superintelligence, russell2015research}.
    \item \textbf{Mitigating Risks of Large-Scale or Long-Term Harms:} Addressing potential downstream effects such as systemic biases, unintended consequences, and even existential risks associated with the deployment of advanced AI systems \cite{amodei2016concrete, brundage2018malicious}.
\end{itemize}
AI safety thus serves as the ethical and operational compass for AI development, ensuring that systems are not only secure but also beneficial and aligned with human welfare.

\section{AI Security as the First Line of Defense}\label{sec:security-first}

The premise of this manuscript is that robust AI security forms the essential groundwork for achieving trustworthy AI systems. By first securing the technical backbone of an AI system, subsequent efforts in AI safety and ethical governance can be more effectively realized.

\subsection{The Importance of a Security-First Approach}
Prioritizing \textbf{AI security} is indispensable for several critical reasons:
\begin{itemize}
    \item \textbf{Data Protection:}  
    Securing the data pipeline is crucial to prevent unauthorized access and manipulation. Data integrity forms the bedrock of model training; if corrupted or poisoned data infiltrates the system, the resulting model will inherently be unreliable \cite{biggio2012poisoning, steinhardt2017certified}.
    \item \textbf{Model Integrity:}  
    Ensuring that the model’s architecture and learned parameters remain uncompromised protects against adversarial manipulations. Attacks such as adversarial examples or model inversion can undermine even the most sophisticated models if their integrity is not maintained \cite{goodfellow2015explaining, fredrikson2015model}.
    \item \textbf{Pipeline Robustness:}  
    A secure operational pipeline ensures that the entire lifecycle—from data collection and preprocessing to model training, deployment, and post-deployment monitoring—is resilient to attacks. This holistic view of security reduces the risk of systemic failures that might arise from isolated vulnerabilities \cite{zhang2018securing, chen2018detecting}.
\end{itemize}

Adopting a security-first approach implies that security measures are integrated at every stage of system design and deployment. This preemptive stance is vital for several reasons:
\begin{itemize}
    \item \textbf{Cascading Failures:}  
    In complex systems, a breach in one component can quickly propagate, undermining the overall system integrity. A secure foundation mitigates the risk of such cascading failures.
    \item \textbf{Adaptive Threats:}  
    As adversaries continually evolve their methods, a security-first strategy allows for continuous monitoring and dynamic updating of defenses \cite{tallam2025cybersentinel}. This adaptability is key to countering new types of attacks as they emerge.
    \item \textbf{Enabling Trust and Transparency:}  
    Only when the underlying infrastructure is secure can efforts in transparency, accountability, and ethical alignment be reliably implemented. Users and regulators are more likely to trust AI systems that demonstrate robust security practices.
\end{itemize}

A security-first approach is not merely about adding layers of protection after the fact. It requires a paradigm shift where security is embedded in the design, development, and deployment processes from the very beginning. By establishing a strong security baseline, organizations create a resilient platform upon which comprehensive AI safety and ethical governance frameworks can be built, ensuring that AI systems perform reliably even in adversarial environments.

\subsection{Metric-Based Insights}
A critical component of \textbf{AI security} is the development and use of metrics that provide insights into system vulnerabilities. These metrics offer a quantitative foundation for monitoring system behavior, assessing risk levels, and informing proactive defense strategies. The following categories of metrics are particularly significant:

\begin{itemize}
    \item \textbf{Anomaly Detection Metrics:}  
    These metrics are designed to identify deviations from established patterns in data or model outputs. Techniques such as autoencoder-based reconstruction errors \cite{sakurada2014anomaly} and statistical methods (e.g., z-scores, Mahalanobis distance) are commonly used to detect outliers that may indicate data poisoning, model drift, or ongoing adversarial activity. By continuously monitoring these metrics, one can flag abnormal events that might otherwise go unnoticed in complex AI pipelines.

    \item \textbf{Vulnerability Scores:}  
    Vulnerability scores quantify the susceptibility of an AI system to various attack vectors. These scores can be derived by evaluating the sensitivity of a model’s outputs to small input perturbations \cite{goodfellow2015explaining, carlini2017towards}, or by assessing the exposure of different components in the data pipeline. Such scores provide a systematic way to rank potential weaknesses, thereby helping prioritize security improvements. For instance, higher gradient norms might indicate a model that is more vulnerable to adversarial perturbations, while metrics on data diversity and integrity can signal risks in the data acquisition process.

    \item \textbf{Resilience Metrics:}  
    Resilience metrics measure how well an AI system can sustain its performance in the face of adversarial conditions. These are often determined through robustness testing, where the system is subjected to simulated attacks, adversarial examples, or stress tests \cite{madry2018towards, tramer2017ensemble}. Key resilience metrics include the percentage drop in accuracy under adversarial conditions, recovery time following an attack, and the proportion of successful defenses. Such metrics not only reflect the current robustness of the system but also serve as benchmarks for iterative improvements and adaptive defense strategies.
\end{itemize}

Together, these metric-based insights enable continuous monitoring and proactive defense. By establishing a baseline for normal operations and systematically quantifying deviations, organizations can detect vulnerabilities early, evaluate the effectiveness of their defenses, and update their security protocols dynamically to counter evolving adversarial strategies.

\section{Defensive Approaches and Best Practices}\label{sec:defenses}

\subsection{Defensive Techniques}
Mitigating adversarial threats in AI systems requires a multifaceted approach. Over the years, researchers have proposed several defensive strategies that target different aspects of vulnerability \citep{tallam2025odi,tallam2025alignment}. These approaches are not mutually exclusive and are often combined to achieve more robust protection. Key techniques include:

\begin{itemize}
    \item \textbf{Adversarial Training:}  
    This method involves augmenting the training data with adversarial examples, inputs intentionally perturbed to fool the model, to improve the model's robustness. By exposing the model to these challenging examples during training, the system learns to recognize and resist similar attacks during deployment \cite{madry2018towards}. Adversarial training has proven effective against a wide range of perturbations \cite{tallam2025immune}, although it often comes at the cost of increased training time and computational resources.
    
    \item \textbf{Differential Privacy:}  
    Differential privacy techniques add carefully calibrated noise to the training process, ensuring that the contribution of any single data point is obscured. This approach protects sensitive individual data points from being reverse-engineered or inferred through model outputs \cite{abadi2016deep}. By balancing utility and privacy, differential privacy not only defends against privacy breaches but can also act as a regularizer, potentially improving generalization.
    
    \item \textbf{Robust Architecture Design:}  
    Some architectures are inherently more resistant to adversarial perturbations. Techniques such as defensive distillation involve training a model to output softened probability distributions over classes, which can reduce the sensitivity to small input changes \cite{papernot2016distillation}. Other architectural innovations include the use of capsule networks or ensemble models that combine predictions from multiple sub-models, thereby reducing the impact of any single point of failure.
    
    \item \textbf{Model Monitoring and Logging:}  
    Continuous monitoring of model inputs, outputs, and internal activations is critical for detecting anomalies that may signal an ongoing attack. Implementing comprehensive logging systems allows practitioners to track unusual patterns and trigger alerts for further investigation. Such systems can also provide valuable data for forensic analysis after an attack, aiding in the refinement of defensive strategies.
\end{itemize}

Collectively, these defensive techniques provide a layered defense strategy, ensuring that even if one line of defense is breached, other safeguards remain in place to mitigate the impact of an adversarial attack.

\subsection{Secure AI Pipelines}
Ensuring the security of an AI system requires a holistic approach that spans the entire lifecycle, from data collection to ongoing maintenance. Each stage of the pipeline presents unique challenges and potential vulnerabilities that must be addressed to ensure overall system resilience:

\begin{enumerate}
    \item \textbf{Data Collection:}  
    The foundation of any AI system is its data. Ensuring data provenance is essential; this involves verifying the source, authenticity, and integrity of the data. Access control mechanisms should be implemented to prevent unauthorized data tampering, and data should be anonymized where possible to mitigate privacy risks. Techniques such as secure multi-party computation and blockchain-based audit trails can further robustify trust in the data acquisition process.
    
    \item \textbf{Model Training:}  
    Secure model training requires the use of trusted and controlled computational environments. This includes using encrypted storage for sensitive data and ensuring that training processes are isolated from external networks to prevent data leakage. Privacy-preserving training protocols, such as federated learning and secure aggregation, enable collaborative model training while maintaining data confidentiality.
    
    \item \textbf{Deployment:}  
    Once trained, models are deployed in environments where they may be exposed to external inputs. During deployment, it is crucial to monitor inference requests for abnormal usage patterns that could indicate a probing or extraction attack. Techniques such as rate-limiting, authentication, and the use of API gateways can help control access and protect the model from abuse. Additionally, deploying models behind secure, encrypted channels further safeguards against interception and tampering.
    
    \item \textbf{Maintenance:}  
    The threat landscape is continuously evolving, and so too must the security measures protecting an AI system. Regular updates and patches are necessary to address newly discovered vulnerabilities. Maintenance also includes periodic retraining of models with updated data and monitoring logs for signs of persistent or emerging threats. Establishing protocols for incident response and recovery ensures that, in the event of an attack, the system can quickly revert to a secure state.
\end{enumerate}

By addressing security at every stage of the AI lifecycle, organizations can build resilient systems that not only perform effectively under normal conditions but also withstand and recover from adversarial challenges. This comprehensive approach is essential for safeguarding sensitive data, protecting intellectual property, and maintaining user trust in AI-driven applications.

\section{Integrating AI Security with AI Safety}\label{sec:integration}

The convergence of AI security and AI safety is essential for developing trustworthy, resilient systems. While AI safety focuses on ensuring that systems act in accordance with ethical standards and intended goals, AI security lays the technical foundation that protects these systems from external and internal threats. When combined, these two pillars form a robust framework that not only prevents harmful outcomes but also builds stakeholder trust through transparency and accountability.

\subsection{A Complementary Relationship}

Although distinct in their objectives, AI security and AI safety share a fundamentally interdependent relationship. While \textbf{AI safety} focuses on ethical design principles, transparency in decision-making, and alignment with human values, these efforts become vulnerable without robust underlying security \cite{tallam2025moral}. Any security breach, such as data poisoning, adversarial manipulations, or model extraction, can severely compromise safety protocols, potentially leading to harmful and unintended outcomes.

Prioritizing security as the foundational layer significantly augments safety initiatives by:
\begin{itemize}
    \item \textbf{Reducing Risk from Data Breaches and Adversarial Attacks:} Strong security protocols safeguard sensitive data and model parameters from unauthorized access or manipulation, ensuring that ethical decision-making processes remain reliable and trustworthy.
    
    \item \textbf{Improving Transparency and Accountability:} Secure infrastructures preserve the integrity of audit trails, transparency records, and system logs. This ensures accurate and trustworthy evaluations of system behavior, thereby bolstering accountability and fostering stakeholder trust.
    
    \item \textbf{Creating a Robust Foundation for Ethical Governance:} A secure environment provides a stable, resilient platform upon which comprehensive ethical guidelines and safety mechanisms can function effectively, significantly reducing the likelihood of cascading failures or malicious interference.
\end{itemize}

In essence, security is not merely a technical addition but a fundamental prerequisite for effective safety implementation. Ensuring that core systems are secure provides the necessary confidence and stability for meaningful deployment and ongoing maintenance of safety measures, ultimately strengthening the reliability and trustworthiness of AI systems.

\subsection{Case for Prioritizing Security}

Placing \textbf{AI security} at the forefront does not diminish the importance of ethical and safety considerations; instead, it establishes a hierarchical framework in which a secure technical foundation is the pre‑condition for any effective safety protocol. We propose a three‑phase model:

\begin{enumerate}
    \item \textbf{Security First}: The initial line of defense is a hardened infrastructure that protects data, models, and pipelines. This includes applying state‑of‑the‑art techniques -- adversarial training, differential privacy, robust architecture design, and continuous monitoring -- to identify, prevent, and mitigate potential threats before deployment.
    
    \item \textbf{Safety Next}: With a secure substrate in place, organizations can layer on safety mechanisms such as transparent decision‑making, ethical guidelines, and accountability frameworks. A protected core sharply reduces the attack surface, allowing these higher‑level safeguards to function reliably and to detect or contain anomalous behavior more effectively.
    
    \item \textbf{Continuous Improvement}: Security and safety are dynamic properties. Ongoing assessment, driven by metric‑based insights such as anomaly scores, vulnerability indices, and resilience metrics, creates a feedback loop that refines both defenses and safety protocols. As adversarial tactics evolve, the system adapts, preserving robustness and ethical alignment over time.
\end{enumerate}

This security‑first hierarchy catalyzes an integrated, self‑reinforcing posture: a fortified infrastructure enables stronger safety controls, while real‑time security metrics guide iterative refinement. The result is an AI ecosystem that is simultaneously robust and ethically aligned, fostering durable trust among users, regulators, and stakeholders.

\section{Conclusions and Future Directions}

This manuscript advances the thesis that \textbf{AI security} must constitute the foundational layer of any trustworthy AI ecosystem. By embedding robust security controls at every stage of the AI lifecycle, that is, data collection, model training, deployment, and maintenance, organizations can safeguard sensitive data, protect intellectual property, and provide a resilient substrate for higher‑level goals of \textbf{AI safety}, transparency, and accountability. A security‑first posture ensures that even against sophisticated adversarial threats, system integrity and reliability are preserved, thereby fostering durable trust among users, regulators, and other stakeholders.

Our survey of state‑of‑the‑art defenses, including adversarial training, differential privacy, resilient architectures, and continuous monitoring, underscores the value of a layered, adaptive strategy. These techniques not only neutralize current attack vectors but also create a foundation for rapidly evolving defenses as new threats emerge. 

\subsection*{Promising Research Directions}

\begin{itemize}
    \item \textbf{Standardized Security Benchmarks:}  
          Developing openly available, domain‑agnostic benchmarks will enable rigorous, reproducible comparisons of defensive techniques across heterogeneous model architectures and threat scenarios. 
    
    \item \textbf{Zero‑Knowledge Proofs for Model Assurance:}  
          Applying zero‑knowledge proof protocols could allow third‑party validation of model integrity and performance without revealing proprietary parameters, balancing confidentiality with verifiability.
    
    \item \textbf{Continuous Verification Frameworks:}  
          Formalizing real‑time verification pipelines, integrating anomaly detection, vulnerability scoring, and automated remediation, will help ensure security measures remain effective throughout the model lifecycle.
    
    \item \textbf{Adaptive and Autonomous Defense Systems:}  
          Leveraging meta‑learning and online adaptation to predict, detect, and respond to novel attack strategies can yield self‑healing AI systems whose defenses evolve in tandem with the threat landscape.
\end{itemize}

\subsection*{Closing Remarks}

A \emph{security‐first} mindset is not a discretionary enhancement but a non‑negotiable prerequisite for credible, high‑stakes AI deployments.  This paper contributes three core value propositions that reinforce that claim:

\begin{enumerate}[leftmargin=*,labelsep=0.8em]
    \item \textbf{We introduce a clear \emph{hierarchy of AI challenges}} in which security forms the bedrock for safety, transparency, and ethical governance. By disentangling and strategically linking these layers, we provide practitioners with a mental model for resource allocation and risk triage.

    \item \textbf{Beyond taxonomy, we survey state‑of‑the‑art defensive techniques}, map them to concrete threat models, and propose \emph{metric‑driven dashboards} (anomaly scores, vulnerability indices, resilience curves) that enable continuous, evidence‑based security auditing across the AI lifecycle.

    \item \textbf{We outline research priorities}: standardized security benchmarks, zero\nobreakdash-knowledge proof protocols for model assurance, continuous verification pipelines, and adaptive, autonomous defenses, that chart a path toward self‑healing, provably robust AI systems.
\end{enumerate}

Taken together, these contributions transform AI security from a reactive afterthought into an \emph{architectural principle}.  As adversarial capabilities grow more sophisticated, proactive, standardized, and empirically validated security practices will be indispensable.  Integrating those practices with comprehensive safety frameworks ensures that future AI systems are not only performant, but also \emph{trustworthy, resilient, and ethically aligned}.  Realizing this vision will require sustained, cross‑disciplinary collaboration among researchers, industry practitioners, regulators, and policymakers, so that the benefits of AI are widely and safely distributed.

\section{Acknowledgements}
The author thanks work colleagues at SentinelAI (Sentinel.AI), UC Berkeley, Anthropic, Google, DOE, DHS and LBNL for their inspiration and feedback on this manuscript.

\bibliographystyle{plainnat}   % or your preferred style
\bibliography{references}

\end{document}